\documentclass[english,12pt]{article}
\usepackage{array}
\usepackage{graphicx}
\usepackage{amssymb}
\usepackage{amsmath}
\usepackage{multirow}
\usepackage{prettyref}
\usepackage{babel}
\usepackage{units}
\usepackage{babel}
\usepackage{cite}
\def\@fmsl@sh#1#2#3{\m@th\ooalign{$\hfil#1\mkern#2/\hfil$\crcr$#1#3$}}
 \def\eq#1\en{\begin{equation}#1\end{equation}}
\def\s[#1,#2]{[#1\stackrel{\star}{,}#2]}
\def\sx[#1,#2]{[#1\stackrel{\star_{x}}{,}#2]}

\newcommand{\nc}{\newcommand}
\nc{\beq}{\begin{equation}}
\nc{\eeq}{\end{equation}}
\nc{\beqa}{\begin{eqnarray}}
\nc{\eeqa}{\end{eqnarray}}

\def\bc{\begin{center}}
\def\ec{\end{center}}

\def\gsim{\mathrel{\mathpalette\atversim>}}

\def\bc{\begin{center}}
\def\ec{\end{center}}

\def\gsim{\mathrel{\rlap{\lower4pt\hbox{\hskip1pt$\sim$}}

    \raise1pt\hbox{$>$}}}       

\def\gsim{\mathrel{\rlap{\lower4pt\hbox{\hskip1pt$\sim$}}
    \raise1pt\hbox{$>$}}}       

\begin{document}
\makeatletter
\def\fmslash{\@ifnextchar[{\fmsl@sh}{\fmsl@sh[0mu]}}
\def\fmsl@sh[#1]#2{%
  \mathchoice
    {\@fmsl@sh\displaystyle{#1}{#2}}%
    {\@fmsl@sh\textstyle{#1}{#2}}%
    {\@fmsl@sh\scriptstyle{#1}{#2}}%
    {\@fmsl@sh\scriptscriptstyle{#1}{#2}}}
\def\@fmsl@sh#1#2#3{\m@th\ooalign{$\hfil#1\mkern#2/\hfil$\crcr$#1#3$}}
\makeatother

\thispagestyle{empty}
\begin{titlepage}
\boldmath
\begin{center}
  \Large {\bf  Frame Transformations of Gravitational Theories}
    \end{center}
\unboldmath
\vspace{0.2cm}
\begin{center}
{ {\large Xavier Calmet}\footnote{x.calmet@sussex.ac.uk} and {\large Ting-Cheng Yang}\footnote{ty33@sussex.ac.uk}
}
 \end{center}
\begin{center}
{\sl Physics $\&$ Astronomy, 
University of Sussex,   Falmer, Brighton, BN1 9QH, UK 
}
\end{center}
\vspace{5cm}
\begin{abstract}
\noindent
We show how to map gravitational theories formulated in the Jordan frame to the Einstein frame at the quantum field theoretical level considering quantum fields in curved space-time.  As an example, we consider gravitational theories in the Jordan frame of the type $F(\phi,\, R)=f(\phi)\, R-V(\phi)$ and perform the  map  to the Einstein frame.   Our results can easily be extended to any gravitational theory. We consider the Higgs inflation model as an application of our results.
\end{abstract}  
\end{titlepage}



\newpage

\section{Introduction}

General Relativity is an extremely successful theory which has now been probed extensively. The beautifully simple Einstein-Hilbert action
\begin{eqnarray}
S_{EH}= \int\, d^{4}x\sqrt{-g} \frac{R}{16\pi G}
\end{eqnarray}
incorporates all our current knowledge of gravity. In this equation, the symbol $G$ stands for Newton's constant, $g$ the determinant of the metric tensor $g_{\mu \nu}$ and $R$ is the Ricci scalar which is uniquely determined by the metric tensor. However, in general, gravitational theories will contain higher dimensional terms such as $R^2$ or $R_{\mu\nu} R^{\mu\nu} $ and fields of different spins. For example when Einstein's gravity is coupled to the Standard Model one needs to introduce particles of spin 0, 1/2 and 1 on top of the spin two metric tensor which represents the graviton. In inflationary theories, one often introduces an inflaton which is represented by a scalar degree of freedom.  

As we shall see shortly, the coupling of scalar fields  allows for interesting complications as scalar fields can be coupled naturally in a non-minimal way to the gravitational field. With the discovery of a scalar boson at the CERN Large Hadron Collider we now know that there are such elementary scalar fields in nature. For example  a neutral scalar field $\phi$ can be coupled to the Ricci scalar using $\phi^2 R$ which is a dimension four operator. Such a non-minimal coupling leads to an action of the type
\begin{eqnarray}
S_{grav}= \int\, d^{4}x\sqrt{-g}\left (\frac{R}{16\pi G}+\frac{1}{2} \xi \phi^2 R \right )
\end{eqnarray}
where $\xi$ is the non-minimal coupling of the field $\phi$ to the curvature scalar.

A general gravitational theory will contain scalar fields. In these generalized scalar-tensor gravitational theories \cite{Starobinsky:1980te}, the notion of frame becomes relevant. It is common to  construct a gravitational theory using one set of variables and then to map the original fields to  another set of variables to make calculations easier. This is called a field redefinition. Each individual set of variables is conventionally called frame. The Jordan frame is the one in which  the gravitational field couples with scalar fields in a  non-minimal way while the Einstein frame is the one where such non-minimal couplings are absent. In other words,  starting from the Jordan frame,  it is always possible to perform a redefinition of the scalar field and the metric tensor to obtain a Lagrangian without the non-minimal coupling. This frame is called the Einstein frame since the coefficient in front of the Ricci scalar is $1/(16\pi G)$ and thus has the form of the usual Einstein-Hilbert action.  While it is often convenient to build a model in the Jordan frame, calculations may appear to be more difficult to perform using these degrees of freedom and frequently relativists and cosmologists transform their models to the Einstein frame to compare with existing inflationary calculations to bound the parameters of their models.  The transformation between these two frames are governed by certain field reparameterizations relating the scalar field, metric and all the other relevant variables,  see e.g.  \cite{Deser:1983rq,Maeda:1988ab,Faraoni:1998qx,Whitt:1984pd,Magnano:1987zz,Maeda:1987xf,Ferraris:1988zz,Jakubiec:1988ef,Barrow:1988xh,DeFelice:2010aj}. 

This concept can be generalized to any number of scalar fields and arbitrarily complicated couplings between the scalar fields and the metric.  While it is well understood how to map theories formulated in the Jordan frame to the Einstein frame, we show how to perform this map at the quantum level. We emphasize that the transformation from one frame to another only implies a field redefinition of the scalar field and the metric tensor. In a path integral formulation of quantum field theory (see e.g. \cite{Chaichian:2001cz,Chaichian:2001da} for a nice introduction), fields are dummy variables which are summed over. As long as the field redefinition does not violate any of the symmetries of the model,  physics cannot be affected and physics cannot depend on the frame.  For example,  field redefinitions are an important part of the renormalization program \cite{Weinberg:1995mt}. At the classical level we find that there is a boundary term which needs to be matched precisely when comparing a physical process described by a theory in the Einstein or in the Jordan frame. The importance of boundary terms has been noted in \cite{Saltas:2010ga}. At the quantum field theoretical (keeping the metric classical) level we find new non-local terms which need to be taken into account when comparing the different frames.
 
 We shall first review the field redefinition used when transforming a gravitational theory from the Jordan to the Einstein frame at the classical level and then compare the theories at the semi-classical level, i.e., we shall not attempt to quantize gravity and will only consider quantum effects of the scalar field.  
  
\section{Review of the classical equivalence and boundary terms}

Before studying a general  class of $F(R)$ theories, we shall review the case of a scalar field non-minimally coupled to the Ricci scalar. Note that we shall consider the transformation from the Jordan frame to the Einstein frame, but our results can trivially be used to consider the reversed transformation from  the Einstein frame to the  Jordan frame.

\subsection{Non-minimally coupled scalar field}

As a prelude to  the general discussion, we first consider the case of a  non-minimally  coupled scalar field $\frac{1}{2}\xi\phi^{2}R$. We define
\begin{eqnarray}
\Omega^{2}=\,\exp[\sigma(x)]=1-8\pi G\xi\phi^{2}.
\end{eqnarray}
The transformations of the metric and scalar field are given by:
\begin{eqnarray} 
\tilde{g}_{\mu\nu} & = & \Omega^{2}g_{\mu\nu}\label{transf}\\ \nonumber
\sqrt{-\tilde{g}} & = & \Omega^{4}\sqrt{-g}\nonumber\\
d\tilde{\phi} & = & \frac{(1-8\pi G\xi(1-6\xi)\phi^{2})^{\nicefrac{1}{2}}}{1-8\pi G\xi\phi^{2}}d\phi\nonumber\\
\tilde{V}(\tilde{\phi}) & = & \Omega^{-4}V(\phi)\nonumber\\ \nonumber
\tilde{R}  & = & \Omega^{-2}(R-\frac{12\square\sqrt{\Omega}}{\sqrt{\Omega}}-3\frac{g^{ab}\nabla_{a}\Omega\nabla_{b}\Omega}{\Omega^{2}}).
\end{eqnarray}
 These transformations\footnote{The potential term is introduced here for the sake of completeness. The
following discussion focuses mainly on the free theory. Also the potential
term does not contribute to the stress tensor and thus to the Jacobian because
the stress tensor comes from the functional differential with respect
to metric. In addition, the transformation is only singular at $\phi=(8\pi G\xi)^{-\frac{1}{2}}$
resulting in no gravitation degree of freedom and $\phi=0$. We do not consider these two trivial cases.}  are merely field redefinitions. 

We start from the usual Jordan frame action 
\begin{equation} 
S_{J}=\int\, d^{4}x\sqrt{-g}\left ( \left (\frac{1}{16\pi G}-\frac{1}{2}\xi\phi^{2} \right )R+\frac{1}{2}g_{\mu\nu}\nabla^{\mu}\phi\nabla^{\nu}\phi-V(\phi)\right )\label{eq:Jordan Frame action}
\end{equation}
which is equivalent to  
\begin{align}
S_{J}  =\int\, d^{4}x\sqrt{-g} \left (\frac{R}{16\pi G}+\left(-\frac{1}{2}\right)\phi \left (\square+\xi R\right)\phi-V(\phi)+\frac{1}{2}\partial_{\mu}(g^{\mu\nu}\phi\partial_{\nu}\phi) \right)\label{eq:Jordan Frame action-1}
\end{align}
after a partial integration of the kinetic term of the scalar field. Using the inverse of the field redefinitions described in eqs. (\ref{transf}) one obtains the action in the Einstein frame
\begin{eqnarray}
S_{E+boundary}  =  \int\, d^{4}x\sqrt{-\tilde g} \left (\frac{\tilde{R}}{16\pi G} -\frac{1}{2}\,\tilde{\phi}\tilde{\square}\tilde{\phi}-\tilde{V}(\tilde{\phi})  +\frac{1}{2}\partial_{\mu}(\tilde{g}^{\mu\nu}\tilde{\phi}\partial_{\nu}\tilde{\phi})-\frac{3\Omega^{2}\tilde{\square}\ln\Omega}{8\pi G\Omega^{2}} \right).\label{eq:Boundary terms after transformation}
\end{eqnarray}
The last two terms are the boundary terms which in flat space-time would be discarded but which can be crucial in curved space-time. For the sake of simplicity, we introduce the following notation for these terms:  $\mathbf{boundary\, terms}$
in the Lagrangian and $\mathbf{surface\, terms}$ in the action
\begin{align}
 & \int\, d^{4}x\,\sqrt{-\tilde g} \left (\frac{1}{2}\partial_{\mu}(\tilde{g}^{\mu\nu}\tilde{\phi}\partial_{\nu}\tilde{\phi})-\frac{3\Omega^{2}\tilde{\square}\ln\Omega}{8\pi G\Omega^{2}} \right)\nonumber \\
 & =\int\, d^{4}x \sqrt{-\tilde g} \,\partial_{\mu}[\frac{1}{2}\tilde{g}^{\mu\nu}\tilde{\phi}\partial_{\nu}\tilde{\phi}-\frac{3\tilde{g}^{\mu\nu}\partial_{\nu}\ln\Omega}{8\pi G}]\nonumber \\
 & =\int\, d\sigma\frac{1}{2}[\tilde{g}^{\mu\nu}\tilde{\phi}\partial_{\nu}\tilde{\phi}-\frac{3\tilde{g}^{\mu\nu}\partial_{\nu}\ln\Omega}{4\pi G}]\mid_{\partial}\nonumber \\
 & \equiv(\mathbf{surface\, terms}),\label{eq:boundary expression}
\end{align}
where $d\sigma$ is the 3-dimensional volume element. Note that the covariant derivatives can be replaced by ordinary derivatives
as we are dealing with scalar fields. One finds that the actions are equivalent up to some boundary terms: 
\begin{equation}
S_{J}=S_{E}+(\mathbf{surface\, terms})\label{eq:serface terms of S}
\end{equation}
 and 
\begin{equation}
\mathcal{L}_{J}=\mathcal{L}_{E}+\partial_{\mu}(\mathbf{boundary\, terms})\label{eq:boundary terms of L}
\end{equation}
with the understanding that 
\begin{eqnarray}
S_{E} & = & \int\, d^{4}x\sqrt{-\tilde g} \left (\frac{\tilde{R}}{16\pi G} -\frac{1}{2}\,\tilde{\phi}\tilde{\square}\tilde{\phi}-\tilde{V}(\tilde{\phi})  \right).
\end{eqnarray}

Note that we can start in the Jordan frame with non trivial boundary conditions such as Gibbons-Hawking terms if an open space is considered without any complication. 

\subsection{Transformation of the action for $F(R)$ scalar-tensor gravitational theories}

In the following we considered the mapping of a $F(R)= f(\phi)R-V(\phi)$ theory, i.e. in the Jordan frame, to the Einstein frame.  These models represent a subset of the ordinary $F(R)$ gravity models. This case is the generalization
of the previous one. If we take $f(\phi)=\phi^{2}$ we recover  the results
obtained for the non-minimally coupled scalar field.

For the class of theories considered here, the conformal factor is given by 
\begin{eqnarray}
\Omega^{2}=16\pi G\left|\frac{\partial F}{\partial R}\right|
\end{eqnarray}
and one has the following field redefinitions:
\begin{eqnarray}
\tilde{g}_{\mu\nu} & = & \Omega^{2}g_{\mu\nu}\label{transf2}\\
\sqrt{-\tilde{g}} & = & \Omega^{4}\sqrt{-g} \nonumber\\
\tilde{\phi} & = & \frac{1}{\sqrt{8\pi G}}\int\,\left (\frac{2f(\phi)+6(\frac{d\, f}{d\phi})^{2}}{4\, f^{2}(\phi)}\right)^{\nicefrac{1}{2}}d\phi \nonumber\\
\tilde{V}(\tilde{\phi}) & = & \Omega^{-4}V(\phi) \nonumber\\
\tilde{R} 
 & = & \Omega^{-2}(R-\frac{12\square\sqrt{\Omega}}{\sqrt{\Omega}}-3\frac{g^{ab}\nabla_{a}\Omega\nabla_{b}\Omega}{\Omega^{2}}). \nonumber
\end{eqnarray}
The Jordan frame action 
\begin{eqnarray}
S_{J}=\int\, d^{4}x\,\sqrt{-g}\,[F(\phi,\, R)+\frac{1}{2}g_{\mu\nu}\nabla^{\mu}\phi\nabla^{\nu}\phi]
\end{eqnarray}
can be mapped to the Einstein frame for the special case $F(\phi,\, R)=f(\phi)\, R-V(\phi)$
using the field redefinitions given above. One finds
\begin{eqnarray}
S_{E} & = & \int\, d^{4}x\sqrt{-\tilde{g}}\,[-\frac{1}{2}\,\tilde{\phi}\tilde{\square}\tilde{\phi}-U(\tilde{\phi})]+\frac{\tilde{R}}{16\pi G}\nonumber \\
 &  & +\frac{1}{2}\partial_{\mu}(\tilde{g}^{\mu\nu}\tilde{\phi}\partial_{\nu}\tilde{\phi})+\frac{3\tilde{\square}\ln\Omega}{8\pi G},\label{eq:Boundary terms after transformation-1}
\end{eqnarray}
where $U(\tilde{\phi})=[16\pi G\mid f(\phi)\mid]^{-2}V(\phi)$. Note that the boundary terms have the same form as that of \prettyref{eq:Boundary terms after transformation}.
The last two terms lead to the boundary terms:
\begin{align}
 & \int\, d^{4}x\,\frac{1}{2}\partial_{\mu}(\tilde{g}^{\mu\nu}\tilde{\phi}\partial_{\nu}\tilde{\phi})+\frac{3\tilde{\square}\ln\Omega}{8\pi G}\nonumber \\
 & =\int\, d^{4}x\,\partial_{\mu}[\frac{1}{2}\tilde{g}^{\mu\nu}\tilde{\phi}\partial_{\nu}\tilde{\phi}-\frac{3\tilde{g}^{\mu\nu}\partial_{\nu}\ln\Omega}{8\pi G}]\nonumber \\
 & =\int\, d\sigma[\frac{1}{2}\tilde{g}^{\mu\nu}\tilde{\phi}\partial_{\nu}\tilde{\phi}-\frac{3\tilde{g}^{\mu\nu}\partial_{\nu}\ln\Omega}{8\pi G}]\mid_{\partial}\nonumber \\
 & \equiv(\mathbf{surface\, terms})\label{eq:boundary expression-1}
\end{align}
which are identical to those found in  Eqs. \prettyref{eq:serface terms of S} and \prettyref{eq:boundary terms of L} previously.

\section{Frame transformation at the quantum level\label{sec:The-Quantum-Level}}

In this section we consider the path integral quantization formalism. The partition function for the Jordan frame theory is given by:
\begin{eqnarray}
Z_{J} & = & N\,\int\, d\mu[\phi]\,\exp\left(\frac{i}{\hbar}\left (\int\, d^{4}x\mathcal{L}_{J}+\int\, d^{4}x\sqrt{-g}\, J_{\phi}\phi\right) \right).
\end{eqnarray}
We now show that it is equivalent to the partition function of the gravitational theory in the Einstein frame if the field redefinition is done properly. We shall actually work backwards and start from $Z_{E}$ defined by
\begin{eqnarray}
Z_{E} & = & \tilde{N}\,\int\, d\mu[\tilde{\phi}]\,\exp\left(\frac{i}{\hbar}\left(\int\, d^{4}x\mathcal{L}_{E}+\int\, d^{4}x\sqrt{-\tilde{g}}\,\tilde{J_{\phi}}\tilde{\phi}\right)\right)
\end{eqnarray}
 and do  field redefinitions using the prescription defined above. We obtain
\begin{eqnarray} \label{eqJac}
Z_{E}   =  \tilde{N}\,\int\,\det C_{N^{\prime}N}d\mu[\phi]\,\exp\frac{i}{\hbar}\left (\int\, d^{4}x(\mathcal{L}_{J}-\partial_{\mu}(\mathbf{boundary\, terms})) +\sqrt{-g}\,J_{\phi}\phi\right),
\end{eqnarray}
where $ C_{N^{\prime}N}$ is the Jacobian of the transformation of the measure of the path integral. It is defined by
\begin{equation}
d\mu[\tilde{\phi}_{N^{\prime}}]=\det C_{N^{\prime}N}d\mu[\phi_{N}],\label{eq:Jacobian in the measure}
\end{equation}
where the $N$ and $N^{\prime}$ represent the modes of the scalar fields in curved space. In the sequel, we will omit the indices.
One can show that the Jacobian is proportional to the trace of the stress tensor. Note that the calculation is identical to the famous anomaly result \cite{Fujikawa:1980eg,Fujikawa:1980vr,Fujikawa:1979ay}, it is however conceptually very different from the anomaly calculation since we are not considering symmetry transformations but rather field redefintions. One obtains
\begin{equation}
i\hbar\ln(\det C)=\frac{1}{2}\,\int\, dx^{4}\langle T^\mu_{\ \mu}\rangle \label{eq:det C and T}
\end{equation}
with the expectation value of any operator defined in the curved space
\begin{eqnarray}
\langle\mathcal{O}\rangle\equiv\frac{\langle out,0\mid\mathcal{O}\mid out,0\rangle}{\langle out,0\mid in,0\rangle}
\end{eqnarray}
and with the definition of the stress tensor $T_{\mu\nu}= \frac{2}{\sqrt{-g}}\frac{\delta S}{\delta g^{\mu\nu}}$
and $W=-i\ln Z[0]$. This leads to 
\begin{eqnarray}
\langle T^{\mu}_{\ \mu}\rangle & = & \frac{2}{\sqrt{-g(x)}}g^{\mu\nu}\frac{\delta W}{\delta g^{\mu\nu}}\\
 & = & -\frac{\Omega}{\sqrt{-g(x)}}\frac{\delta W}{\delta\Omega}
\end{eqnarray}
The exact form of $\langle T^{\mu}_{\ \mu}\rangle$ is model dependent and can be calculated for specific models. We present some examples in appendix A. Our main result is
\begin{eqnarray}
Z_{E}   =  \tilde{N}\,\int\,d\mu[\phi]\,\exp\frac{i}{\hbar}\left (\int\, d^{4}x \left (\mathcal{L}_{J}- \frac{1}{2} \langle T^{\mu}_{\ \mu}\rangle -\partial_{\mu}(\mathbf{boundary\, terms})\right) +\sqrt{-g}\,J_{\phi}\phi\right).
\end{eqnarray}
While the classical boundary term was known, the  semi-classical correction is new and should be taken into account when performing the field redefinition which maps a gravitational theory formulated in the Einstein frame to the Jordan frame and vice versa. Assuming that the map is done correctly the two formulations are obviously physically equivalent since we are merely dealing with field redefinitions.

Note that if we treated the metric as a quantum field instead of a classical background as we did, we would obtain a new Jacobian in equation (\ref{eqJac}) corresponding to the field redefinition of the metric. However, this Jacobian corresponds to diagrams with closed graviton loops and scalar fields and gravitons as external lines. These diagrams are not renormalizable within quantum general relativity. This is our main motivation to keep a classical background metric.

Our calculation shows that the partition function of the theory defined in the Einstein frame can be mapped to the Jordan frame in a consistent manner. When mapping the quantum field theory defined in curved space-time, one encounters a Jacobian because of  the transformation of field variables. Since the transformations only involve a redefinition of dummy variables, physics cannot be affected.

The actions of the Jordan and Einstein frames are related by field redefinitions and these two actions are thus obviously equivalent. Indeed one always has the freedom to redefine fields in a theory as long as the Jacobian of the transformation is not singular. This constraint on the Jacobian is from time to time incorrectly interpreted as an anomaly free condition. While it is obviously true that the Jacobian is related to the question of the anomaly  \cite{Birrell:1979ip,Drummond:1977dg,Adler:1969gk,Bell:1969ts,AlvarezGaume:1983ig,Witten:1985xe,Duff:1993wm,Capper:1974ic,Deser:1976yx}, the presence of an anomaly is not an obstacle when changing frames.  If one identifies an anomaly in one frame it is just an indication that the corresponding symmetry is broken in any frame.

\section{Application to Higgs inflation model}

We are now ready to apply our findings to the Higgs inflation model \cite{Bezrukov:2007ep} which has received considerable attention in the last few years. In this minimalistic approach, the Higgs boson of the standard model is the inflaton. It requires a non-minimal coupling $\xi$ of the order of $10^4$. The Lagrangian is given by 
\begin{eqnarray}
L_{tot}=L_{SM} + \frac{(M^2- \xi H^\dagger H)}{2} R.
\end{eqnarray}
where $H$ is the Higgs doublet. Using the unitarity gauge in which $H=1/\sqrt{2}(0, h+v)^{\top}$ and the approximations $v h\ll h^2$ and $M^2-\xi v^2/2=M_P^2$ which are valid during  the inflation regime and for $\xi\sim 10^4$, one has
\begin{eqnarray}
S_J= \int d^x \sqrt{-g} \left ( -\frac{M_P^2-\xi h^2}{2} R - \frac{1}{2} h \Box h - \frac{\lambda}{4} (h^2-v^2)^2 \right )
\end{eqnarray}
The standard procedure to confront this model with CMB data consists in transforming the theory which is originally formulated in the Jordan frame to the Einstein frame. We now calculate using Equation  (\ref{eqapp2}) of the appendix the new contribution to the effective action in the Einstein frame which must be taken into account
\begin{eqnarray}
\langle T_{\ \mu}^\mu\rangle&=&\left (\frac{1}{64\pi^{2}}g_{\ \mu}^\mu \right)  \times  \\ \nonumber && \left (m^{2}\left [m^{2}+\left (\xi-\frac{1}{6}\right )R\right ] 
\left [\Psi\left (\frac{3}{2}+\nu \right )+\Psi\left (\frac{3}{2}-\nu \right )-\ln \left (12m^{2}R^{-1} \right ) \right]  \right .
\\ \nonumber && \left .
-m^{2} \left (\xi-\frac{1}{6} \right )R-\frac{1}{18}m^{2}R-\frac{1}{2} \left (\xi-\frac{1}{6} \right )^{2}R^{2}+\frac{1}{2160}R^{2} \right )
 \end{eqnarray}
where $\Psi(z)=\,\frac{\Gamma^{\prime}(z)}{\Gamma(z)}$, $\nu=\sqrt{\frac{9}{4}-m^{2}\frac{12}{R}-12\xi}$ and $m=\sqrt{\lambda} v/2$.
This is a new term which must be taken into account when considering the Higgs inflation model in the Einstein frame. Note that this is the leading order correction.

\section{Conclusion}
We have shown how to map  gravitational theories formulated in the Jordan frame to the Einstein frame at the quantum field theoretical level.  In a quantum field theory, fields are dummy variables. This is particularly obvious when using the path integral quantization formulation. Field redefinitions cannot affect the calculation of observables. In this paper we consider as an example gravitational theories in the Jordan frame of the type $F(\phi,\, R)=f(\phi)\, R-V(\phi)$ and show that these models can be mapped to the Einstein frame. While it had been known that the theories were equivalent up to a boundary term at the classical level, we show that there is a new term which appears when considering quantum fields in a curved space-time. The physical equivalence of the frames is obvious if the field redefinitions are done properly.  Our results can easily be extended to any gravitational theory. In conclusion, we reaffirm that the frame transformation cannot affect the calculations of observables as long as the proper boundary terms  and Jacobian terms are taken into account. Finally, we provide an application of our results by calculating the leader order term of the effective action needed when considering the Higgs inflation model in the Einstein frame.

\bigskip

\section*{Acknowledgments}

This work is supported in part by the European Cooperation in Science and Technology (COST) action MP0905 ``Black Holes in a Violent  Universe". The work of X.C. was supported by the STFC grant ST/J000477/1.

\section*{Appendix A}

Let us first consider the conformal coupling and take a non-minimal coupling $\xi=\frac{1}{6}$ and a massless case for the scalar field.  Using
the adiabatic expansion, one finds  the regularized expectation value of the energy-momentum tensor
\cite{Dowker:1976zf,Deser:1976yx,Brown:1976wc,Duncan:1976pv,Brown:1977pq,Tsao:1977tj,Wald:1978pj,Birrell:1982ix,parker2009quantum}
\begin{eqnarray}
\langle T_{\ \mu}^{\mu}\rangle&=&\frac{1}{4\pi^{2}}\left [\frac{1}{120}C_{\alpha\beta\gamma\delta}C^{\alpha\beta\gamma\delta}-\frac{1}{360}G+\frac{1}{180}\square R \right ]
\end{eqnarray}
where
\begin{eqnarray}
G = R_{\alpha\beta\gamma\delta}R^{\alpha\beta\gamma\delta}-4R_{\alpha\beta}R^{\alpha\beta}+R^{2}
\end{eqnarray}
is the Gauss-Bonnet topological invariant and 
\begin{eqnarray}
C_{\alpha\beta\gamma\delta}C^{\alpha\beta\gamma\delta}=R_{\alpha\beta\gamma\delta}R^{\alpha\beta\gamma\delta}-2R_{\alpha\beta}R^{\alpha\beta}+\frac{1}{3}R^{2}
\end{eqnarray}
is the square of  the Weyl tensor. The expectation value is purely local
and is only dependent on the geometry and it does not depend on the choice of the vacuum. 

For a general non-minimal coupling, with a massive scalar field $\phi$, the
stress tensor is given by
\begin{eqnarray} 
 T_{\mu\nu}&=&\frac{2}{\sqrt{-g}}\frac{\delta S_\phi}{\delta g^{\mu\nu}} \\ \nonumber
 &=&(1-2\xi)\nabla_{\mu}\phi\nabla_{\nu}\phi+\left (2\xi-\frac{1}{2}\right )g_{\mu\nu}g^{\rho\sigma}\nabla_{\rho}\phi\nabla_{\sigma}\phi \\ \nonumber &&
 +\frac{1}{2}\xi g_{\mu\nu}\phi\square\phi-\xi\left [R_{\mu\nu}-\frac{1}{2}Rg_{\mu\nu}+\frac{3}{2}\xi Rg_{\mu\nu}\right ]\phi^{2}-m^{2}g_{\mu\nu}\phi^{2}.
 \end{eqnarray}
Its expectation value can be evaluated using the adiabatic regularization
scheme in curved space \cite{Dowker:1975tf,Parker:1974qw,Fulling:1974pu,Huang:1991zp}
\begin{eqnarray} \label{eqapp2}
\langle T_{\mu\nu}\rangle&=&(\frac{1}{64\pi^{2}}g_{\mu\nu})(m^{2}[m^{2}+(\xi-\frac{1}{6})R][\Psi(\frac{3}{2}+\nu)+\Psi(\frac{3}{2}-\nu)-\ln(12m^{2}R^{-1})]
\\ \nonumber &&
-m^{2}(\xi-\frac{1}{6})R-\frac{1}{18}m^{2}R-\frac{1}{2}(\xi-\frac{1}{6})^{2}R^{2}+\frac{1}{2160}R^{2})
 \end{eqnarray}
where $\Psi(z)=\,\frac{\Gamma^{\prime}(z)}{\Gamma(z)}$. The contraction
of indices can be worked out in a  straightforward manner.  $\nu$ in  $\Psi(\frac{3}{2}-\nu)$ is defined as follows:
\begin{eqnarray}
[\nu(n)]^{2}=\frac{1}{4}(n-1)^{2}-m^{2}\alpha^{2}-\xi n(n-1)
\end{eqnarray}
 where $n$ is the dimension of space-time and $\alpha$is the radius
of de Sitter universe, which is given by  $R=12\alpha^{-2}$. Thus
in 4 dimension, one has
\begin{eqnarray}
\nu=\sqrt{\frac{9}{4}-m^{2}\frac{12}{R}-12\xi}.
 \end{eqnarray}
Note that $\langle T_{\mu\nu}\rangle$ is dependent not only on the geometry but also on the global (long distance behavior)
and the chosen vacuum  state. This is also dependent on the chosen  background geometry; here we took de Sitter space. Any other $F(R)$ type theory can be treated the same way.

The expectation value of the stress tensor is finite in  flat space
and can be removed via a renormalization of the zero point of the potential energy.
However in the curved space this quantity is
divergent and will need to be renormalized using a renormalization scheme such as  in adiabatic regularization, point splitting, dimensional regularization or the 
$\zeta$ function regularization. It cannot be discarded in curved space-time as one cannot randomly change the ground state energy in the presence of curvature.


\bigskip{}


\baselineskip=1.6pt 
\begin{thebibliography}{10}





\bibitem{Starobinsky:1980te} 
  A.~A.~Starobinsky,
  Phys.\ Lett.\ B {\bf 91}, 99 (1980).  

\bibitem{Faraoni:1998qx} 
  V.~Faraoni, E.~Gunzig and P.~Nardone,
  Fund.\ Cosmic Phys.\  {\bf 20}, 121 (1999)  [gr-qc/9811047].  

\bibitem{Maeda:1988ab} 
  K.~-i.~Maeda,
  Phys.\ Rev.\ D {\bf 39}, 3159 (1989).  




\bibitem{Barrow:1988xh} 
  J.~D.~Barrow and S.~Cotsakis,
   Phys.\ Lett.\ B {\bf 214}, 515 (1988).  


\bibitem{DeFelice:2010aj} 
  A.~De Felice and S.~Tsujikawa,
   Living Rev.\ Rel.\  {\bf 13}, 3 (2010)  [arXiv:1002.4928 [gr-qc]].  
   
 

\bibitem{Deser:1983rq} 
  S.~Deser,
  Phys.\ Lett.\ B {\bf 134}, 419 (1984).  


\bibitem{Ferraris:1988zz} 
  M.~Ferraris, M.~Francaviglia and G.~Magnano,
   Class.\ Quant.\ Grav.\  {\bf 5}, L95 (1988).  


\bibitem{Jakubiec:1988ef} 
  A.~Jakubiec and J.~Kijowski,
   Phys.\ Rev.\ D {\bf 37}, 1406 (1988).  

\bibitem{Maeda:1987xf} 
  K.~-i.~Maeda,
  Phys.\ Rev.\ D {\bf 37}, 858 (1988).  

\bibitem{Magnano:1987zz} 
  G.~Magnano, M.~Ferraris and M.~Francaviglia,
   Gen.\ Rel.\ Grav.\  {\bf 19}, 465 (1987).  

\bibitem{Whitt:1984pd} 
  B.~Whitt,
    Phys.\ Lett.\ B {\bf 145}, 176 (1984).  

\bibitem{Chaichian:2001cz} 
  M.~Chaichian and A.~Demichev,
  ``Path integrals in physics. Vol. 1: Stochastic processes and quantum mechanics,''
  Bristol, UK: IOP (2001) 336 p
  
\bibitem{Chaichian:2001da} 
  M.~Chaichian and A.~Demichev,
  ``Path integrals in physics. Vol. 2: Quantum field theory, statistical physics and other modern applications,''
  Bristol, UK: IOP (2001) 345 p

\bibitem{Weinberg:1995mt} 
  S.~Weinberg,
  ``The Quantum theory of fields. Vol. 1: Foundations,''
  Cambridge, UK: Univ. Pr. (1995) 609 p;
  ``The quantum theory of fields. Vol. 2: Modern applications,''
  Cambridge, UK: Univ. Pr. (1996) 489 p
  
\bibitem{Saltas:2010ga} 
  I.~D.~Saltas and M.~Hindmarsh,
  Class.\ Quant.\ Grav.\  {\bf 28}, 035002 (2011)  [arXiv:1002.1710 [gr-qc]].  

\bibitem{Fujikawa:1980eg} 
  K.~Fujikawa,
  Phys.\ Rev.\ D {\bf 21}, 2848 (1980)  [Erratum-ibid.\ D {\bf 22}, 1499 (1980)].  

\bibitem{Fujikawa:1980vr} 
  K.~Fujikawa,
  Phys.\ Rev.\ Lett.\  {\bf 44}, 1733 (1980).  

\bibitem{Fujikawa:1979ay} 
  K.~Fujikawa,
  Phys.\ Rev.\ Lett.\  {\bf 42}, 1195 (1979).  







\bibitem{Bell:1969ts} 
  J.~S.~Bell and R.~Jackiw,
   Nuovo Cim.\ A {\bf 60}, 47 (1969).  


\bibitem{Birrell:1979ip} 
  N.~D.~Birrell and P.~C.~W.~Davies,
  Phys.\ Rev.\ D {\bf 22}, 322 (1980).  

\bibitem{Capper:1974ic} 
  D.~M.~Capper and M.~J.~Duff,
  Nuovo Cim.\ A {\bf 23}, 173 (1974).  



\bibitem{Deser:1976yx} 
  S.~Deser, M.~J.~Duff and C.~J.~Isham,
   Nucl.\ Phys.\ B {\bf 111}, 45 (1976).  



\bibitem{Drummond:1977dg} 
  I.~T.~Drummond and G.~M.~Shore,
  Phys.\ Rev.\ D {\bf 19}, 1134 (1979).  



\bibitem{Duff:1993wm} 
  M.~J.~Duff,
  Class.\ Quant.\ Grav.\  {\bf 11}, 1387 (1994)  [hep-th/9308075].  



\bibitem{Witten:1985xe} 
  E.~Witten,
  Commun.\ Math.\ Phys.\  {\bf 100}, 197 (1985).  




\bibitem{Adler:1969gk} 
  S.~L.~Adler,
  Phys.\ Rev.\  {\bf 177}, 2426 (1969).  




\bibitem{AlvarezGaume:1983ig} 
  L.~Alvarez-Gaume and E.~Witten,
  Nucl.\ Phys.\ B {\bf 234}, 269 (1984).  


\bibitem{Callan:1970ze} 
  C.~G.~Callan, Jr., S.~R.~Coleman and R.~Jackiw,
  Annals Phys.\  {\bf 59}, 42 (1970).  

\bibitem{Birrell:1982ix} 
  N.~D.~Birrell and P.~C.~W.~Davies,
  ``Quantum Fields In Curved Space,''  Cambridge, Uk: Univ. Pr. ( 1982) 340p



\bibitem{Brown:1976wc} 
  L.~S.~Brown,
  Phys.\ Rev.\ D {\bf 15}, 1469 (1977).  

\bibitem{Brown:1977pq} 
  L.~S.~Brown and J.~P.~Cassidy,
  Phys.\ Rev.\ D {\bf 15}, 2810 (1977).  




\bibitem{Dowker:1976zf} 
  J.~S.~Dowker and R.~Critchley,
   Phys.\ Rev.\ D {\bf 16}, 3390 (1977).  




\bibitem{Duncan:1976pv} 
  A.~Duncan,
    Phys.\ Lett.\ B {\bf 66}, 170 (1977).  



\bibitem{parker2009quantum} 
  L.~Parker and D.~Toms, 
  ``Quantum Field Theory in Curved Spacetime: Quantized Fields and Gravity,''  Cambridge, Uk: Univ. Pr. ( 2009) 472p




\bibitem{Tsao:1977tj} 
  H.~-S.~Tsao,
    Phys.\ Lett.\ B {\bf 68}, 79 (1977).  




\bibitem{Wald:1978pj} 
  R.~M.~Wald,
  Phys.\ Rev.\ D {\bf 17}, 1477 (1978).  




\bibitem{Dowker:1975tf} 
  J.~S.~Dowker and R.~Critchley,
   Phys.\ Rev.\ D {\bf 13}, 3224 (1976).  



\bibitem{Fulling:1974pu} 
  S.~A.~Fulling, L.~Parker and B.~L.~Hu,
    Phys.\ Rev.\ D {\bf 10}, 3905 (1974).  




\bibitem{Huang:1991zp} 
  W.~- H.~Huang,
   Phys.\ Rev.\ D {\bf 43}, 1262 (1991).  



\bibitem{Parker:1974qw} 
  L.~Parker and S.~A.~Fulling,
   Phys.\ Rev.\ D {\bf 9}, 341 (1974).  



\bibitem{Bezrukov:2007ep} 
  F.~L.~Bezrukov and M.~Shaposhnikov,
  Phys.\ Lett.\ B {\bf 659}, 703 (2008)
  [arXiv:0710.3755 [hep-th]].









  
\end{thebibliography}
\end{document}